\documentclass[aps,pre,twocolumn,superscriptaddress,showpacs
]{revtex4-1}

\usepackage{amsmath,amssymb,bm}
\usepackage{graphics}
\usepackage{graphicx}
\usepackage{comment}
\usepackage{mathtools}
\begin{document}
\title{
Ranking scientists
}

\author{S.~N. Dorogovtsev}
\affiliation{Departamento de F{\'\i}sica da Universidade de Aveiro $\&$ I3N, Campus Universit\'ario de Santiago, 3810-193 Aveiro, Portugal}
\affiliation{A.~F. Ioffe Physico-Technical Institute, 194021 St. Petersburg, Russia}
\author{J.~F.~F. Mendes}
\affiliation{Departamento de F{\'\i}sica da Universidade de Aveiro $\&$ I3N, Campus Universit\'ario de Santiago, 3810-193 Aveiro, Portugal}
\begin{abstract} 

{\bf 
Currently the ranking of scientists is based on the $h$-index, which is widely perceived as an imprecise and simplistic though still useful metric. We find that the $h$-index actually favours modestly performing researchers and propose a simple criterion for proper ranking.
}
\end{abstract}

\maketitle

In this age of big data and high social and professional mobility, ranking has become one of the central issues in social life and information technologies.  
Ranking 
algorithms, including the famed Google PageRank \cite{page1999pagerank,brin1998anatomy,brin1998can}, 
enable 
automated 
selection of relevant information and 
the 
functioning of search engines. 
Ranking is an obligatory task of various selection and
evaluation boards. 
At the same time, ranking algorithms are powerful tools for control and manipulation of economies and society,   
allowing quick redirection of web traffic through small biased updates, and restriction of public access to undesired information and the products and services of economic and political rivals. The biased ranking of scientists 
can be particularly harmful, substituting misleading citation-based targets for 
the real aims of scientific research: strong results. One of the consequences of this substitution---a marked reshaping of research behaviors---has already been noticed by sociologists \cite{abbott2010metrics}.

\subsection*{Major metric of scientific productivity}
\label{ss1}

Currently the ranking of scientists is 
largely 
based on  
J.~E.~Hirsch's $h$-index 
to measure ``an individual's scientific research output'' 
\cite{hirsch2005index} 
(note that physicists have 
been keen to study citation statistics  
\cite{redner1998popular,radicchi2008universality,petersen2011statistical,mazloumian2011citation,penner2013predictability}).  
If 
a researcher's papers 
are ranked 
in descending order by 
number of times cited, $c_r$, where 
$r=1,2,\ldots,N$, then the Hirsch index $h$ is the maximum 
$r$ such that $c_r \geq r$. 
To find the $h$-index, we do not need the complete citation record of a scientist, an advantage over total number of citations---the previous main measure of scientific productivity. 
Unfortunately, while it reasonably ignores the tail of the citation record, the major drawback of the $h$-index is that it neglects a researcher's most cited works 
\cite{egghe2006theory,costas2007h}. 
Overall, the $h$-index is widely perceived as an imprecise and simplistic but nonetheless useful and sufficiently reliable metric of scientific productivity \cite{hirsch2007does}. 
Moreover, there is a 
tendency to evaluate a scientist's achievements based exceptionally on this single number. 
Given this total
acceptance, it is surprising that in practically any physics department, you can
find at least one prolific researcher with $h$-index greater than those not only
of famous physicists from previous generations with few papers published but
even of certainly successful and very active scientists of recent years, like, say,
A. M. Polyakov or R. B. Laughlin. Remarkably, each of these quite ordinary
researchers with high $h$-indices typically receives fewer citations in total than
Polyakov or Laughlin, while having a much longer list of publications. 
To see whether the $h$-index indeed systematically favours long lists of modestly cited publications, we
study a representative sample of 
researchers from physics and complex systems
extracted from the Thomson Reuters Web of Science (Core Collection) database (http://apps.webofknowledge.com) and analyze correlations
between their $h$-index and number of papers, $N$, and total number of citations,
$C$.

\begin{figure}[t]
\begin{center}
\includegraphics[scale=0.229]{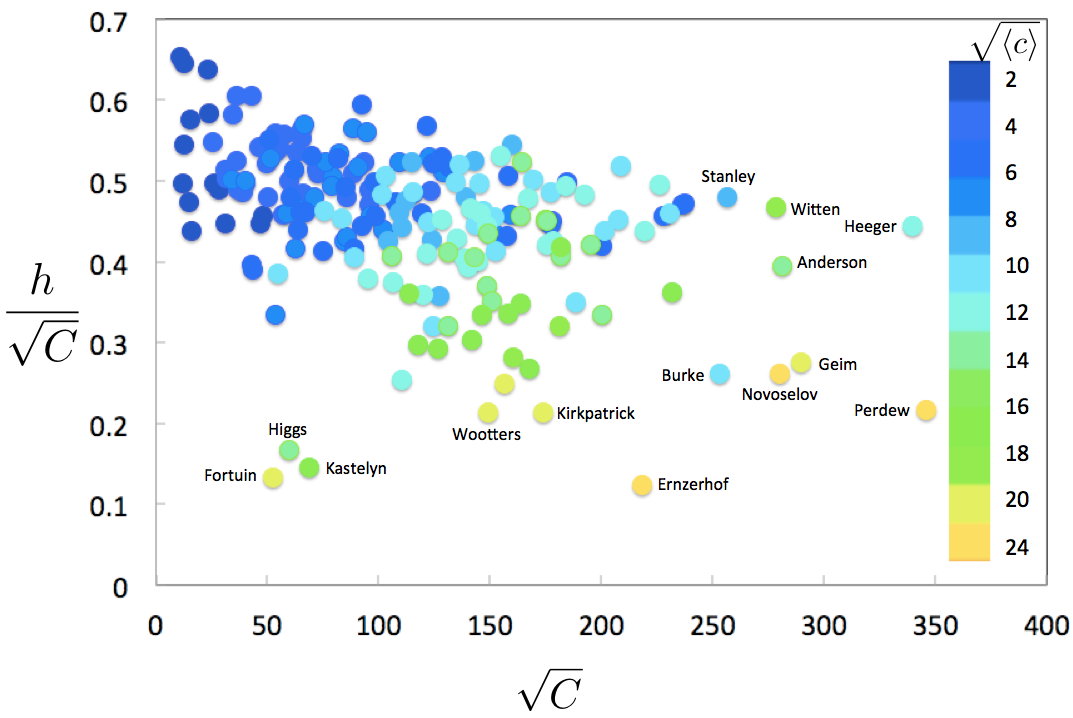}
\end{center}
\caption{Plot of $h/\sqrt{C}$ vs. $\sqrt{C}$ for a 
sample of researchers from physics and complex systems, extracted from the Web of Science database. Here $C$ is the total number of citations to the papers of a researcher, and $h$ is his or her $h$-index. 
$\sqrt{C}$ is the maximum possible value of $h$. 
The color of a dot 
shows the mean number of citations $\langle c \rangle$ per paper for the researcher. The bright dots tend to occur at the bottom of the plot, indicating that for a given $C$, on average, the $h$-index decreases with increasing $\langle c \rangle$. 
In other words, for two researchers receiving the same total number of citations, the one with higher number of publications has, on average, a higher $h$-index. So the $h$-index actually 
punishes 
strong researchers. 
We indicate the scientists names for the points standing out from the crowd, and one can see that the region of small $h/\sqrt{C}$ is completely occupied by outstanding researchers.
}
\label{fi3}
\end{figure}

\subsection*{$h$-index versus the quality of research}
\label{ss2}

Our sample contains 208 scientists with various citation records, at different stage of their academic career and those who have already finished their research activity. 
We include in our sample 
many highly cited researchers, enabling us to measure correlations in a wide range of citation data. 
Typically, citation datasets are highly heterogeneous (in particular, it is difficult to separate science administrators and active researchers), so a larger sample would not improve the statistics. 
Each dot in Fig.~\ref{fi3} shows how the ratio of $h$ and its maximum possible value $\sqrt{C}$ of one of these scientists relates to $\sqrt{C}$. 
The color of a dot indicates the mean number of citations $\langle c \rangle = C/N$ per paper for the researcher. 

One can see that the bright dots tend to occur at the bottom of the plot, indicating that for a given $C$, on average, the $h$-index decreases with increasing $\langle c \rangle$. 
We observe this marked trend everywhere on the plot except the region of low $C$, in which the $h$-index strongly fluctuates. We indicate the scientists names for some points in the figure, and it is easy to see that the region of small ratios $h/\sqrt{C}$, i.e. of the worst  possible $h$, is completely occupied by outstanding researchers. This was already noticed by S.~Redner, who studied the distribution of $h/\sqrt{C}$ \cite{redner2010meaning}.

To quantify this negative gradient we apply the method of least squares to this dataset, which provides the approximation 
$$
0.584 + 0.00023 \sqrt{C} - 0.020 \sqrt{\langle c \rangle}
$$ 
for $h/\sqrt{C}$. We find that the standard deviation is relatively small, $0.057$, compared to the standard deviation $0.091$ from the average value $\langle h/\sqrt{C} \rangle=0.452$ for our sample, so we can rely on this fit. The negative sign of the third term in the expression above confirms that for two researchers receiving the same total number of citations, the one with higher number of publications has, on average, a higher $h$-index. Consequently the Hirsch index is not merely imperfect but it unfairly favours
modestly performing scientists and punishes stronger researchers with a large mean
number of citations per paper.  

\begin{figure}[t]
\begin{center}
\includegraphics[scale=0.235]{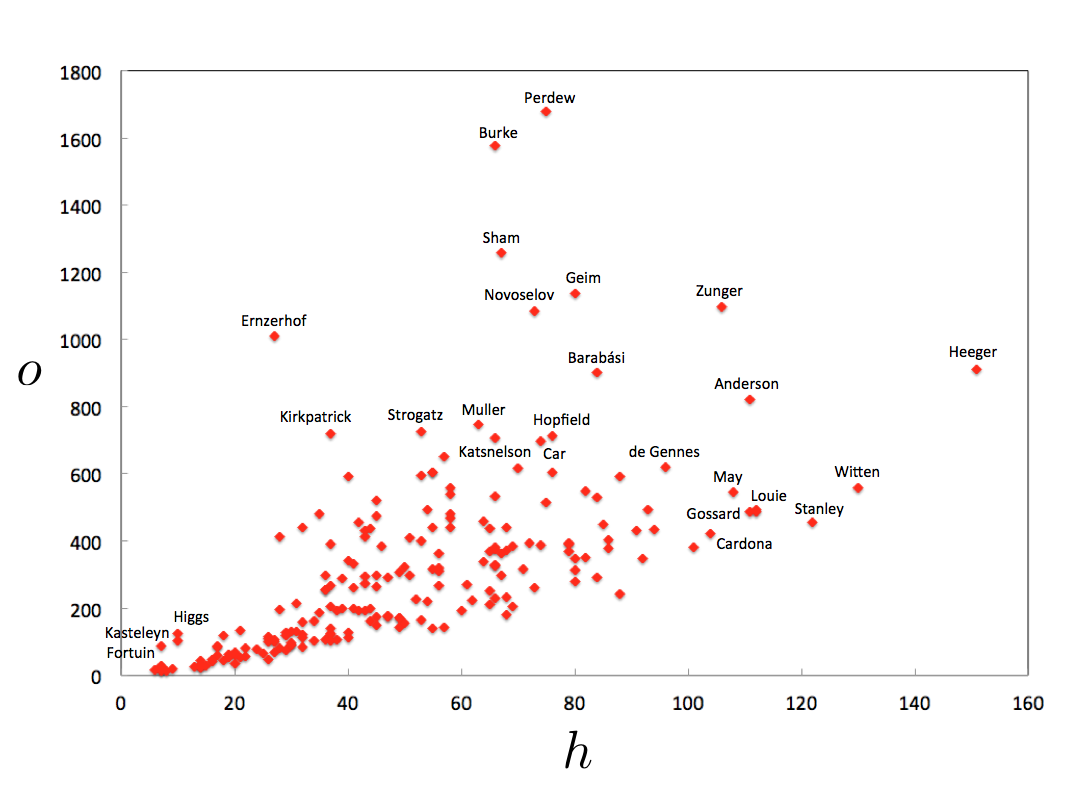}
\end{center}
\caption{Plot of $o$ vs. $h$ for the same sample of scientists as in Fig.~\ref{fi3}. Here $o =
\sqrt{mh}$, where $m$ is the number of citations 
to a researcher's most cited work. 
Projecting the points to the $h$-axis, we get the $h$-index-based ranking, while projecting them to  the $o$-axis, we get the $o$-based ranking. 
We indicate the scientists names for the points standing out from the crowd in this figure. These names strongly overlap with those shown in Fig.~\ref{fi3}. One can see that the $h$-index does not show much favour to A.~K.~Geim and K.~S.~Novoselov, the physicists with recent outstanding achievements, and they are deeply hidden in the $h$-based ranking. In contrast, Geim and Novoselov are among the top physicists according to the $o$-index. At the same time, even the $o$-index cannot move many outstanding scientists of earlier generations to the top of the rating, see P.~W.~Higgs.
}
\label{fi5}
\end{figure}

\subsection*{How to measure research output}
\label{ss3}

The reason why the $h$-index was accepted so immediately and totally is not a question of applicability and relevance but rather of human nature and psychology. Playing with the ranked list of citations to your 
own works is fun. To get your $h$-index, you do not need to make any summations or other ``serious computations''. You only compare two numbers: the number of citations a paper receives and its rank, and there is something particularly attractive in the point where these two numbers meet each other. 
Ironically, the negative feature, namely, the possibility to manipulate and increase your $h$-index by self-citing, only adds to its charm. 
The more you publish, the easier it is to increase the $h$-index in this way. 

So if the $h$-index does not provide a fair ranking, is it at all possible to rank scientists based on a simple metric?
Numerous efforts to introduce a more reasonable measure by trying to include
more detail 
sacrifice the charming simplicity and attractiveness of the Hirsch
index, becoming impractical, yet still fail to achieve their goal \cite{egghe2006theory,tol2009h,jin2007r}. 
Instead, 
we propose to take into account psychology, keeping the elegance and simplicity of the $h$-index. 
We also note that, similarly to the Google PageRank, 
a practical ranking criterion in principle cannot be absolutely precise. 
The idea is to introduce a simple new measure of scientific research output that focuses on a
researcher's most cited paper to substantially indicate his or her major achievement,
good luck, and scientific level, but also accounts for the $h$-index. 

Any researcher remembers the two numbers: the number of citations to his or her most cited paper, $m$, and the $h$-index. We introduce the $o$-index, which is the
geometric mean of 
$m$ and $h$: 
$o = \sqrt{mh}$. Here $m$ accounts for the 
great result, and $h$ accounts for persistence and diligence. The geometric mean is the simplest combination that can be written without knowing the statistics of citations for an individual researcher. 

The $h$- and $o$-indices generate the markedly different rankings. 
Figure~\ref{fi5} for our sample of scientists shows 
how many successful researchers, deeply hidden in the $h$-based ranking, become well visible if we apply the $o$-index. 
According to the $h$-ranking, the top physicist is A.~J.~Heeger, while in the $o$-ranking,  
the first is J.~P.~Perdew. 
These researchers received almost the same number of citations, but Heeger published four times more papers ($1284$) than Perdew ($317$). 
The most cited papers of Heeger and Perdew received $5482$ and $37641$ citations respectively. 
Interestingly, the highest $o$-index for a physicist 
(J.~P.~Perdew, $o=1680$) is not much below the maximum found number for a scientist 
(B.~Vogelstein, medicine, $o=2071$).  

Let us roughly estimate the $o$-index ignoring fluctuations to find how it depends on the number of publications. The estimate of the $h$-index is known: $h \sim \sqrt{C}$ \cite{hirsch2005index,redner2010meaning}. 
The number of citations to the most cited paper falls somewhere between two numbers: 
$
C/N \leq m \leq C$, so, without knowing the statistics of citations for a researcher, we roughly 
estimate it as $m \sim C/\sqrt{N}$. As a result we readily obtain 
$$
o \sim C^{3/4}N^{-1/4} 
\sim C^{1/2}\langle c \rangle^{1/4}.
$$ 
This shows that the $o$-index, on average, grows with increasing mean number of citations per paper as is reasonable for proper ranking. 
For our sample of scientists we find the coefficient $0.88$ of the right-hand-side terms of this relation and the relatively small standard deviation $0.25$, so this simple estimate works 
well. 

It is more difficult to manipulate the $o$-index than the Hirsch one but it is still possible. 
For many, this feature 
may be psychologically attractive. 
Focusing on a most cited paper adds extra interest to citation records. 
At the same time, the $o$-index clearly distinguishes successful
researchers and provides a natural, easily implementable ranking criterion for
scientists. 

The merit of a researcher is determined by his or her strongest results, 
not by the number of publications. We find that the widely used $h$-index-based ranking of scientists consistently contradicts this principle. This is all the more surprising given that one can so easily remove this contradiction and rank scientists reasonably and 
fairly.


\end{document}